# A Semantic VSM-Based Recommender System

Hadi Fanaee Tork and Mehran Yazdi

*Abstract*—Online forums enable users to discuss together around various topics. One of the serious problems of these environments is high volume of discussions and thus information overload problem. Unfortunately without considering the users interests, traditional Information Retrieval (IR) techniques are not able to solve the problem. Therefore, employment of a Recommender System (RS) that could suggest favorite's topics of users according to their tastes could increases the dynamism of forum and prevent the users from duplicate posts. In addition, consideration of semantics can be useful for increasing the performance of IR based RS. Our goal is study of impact of ontology and data mining techniques on improving of content-based RS. For this purpose, at first, three type of ontologies will be constructed from the domain corpus with utilization of text mining, Natural Language Processing (NLP) and Wordnet and then they will be used as an input in two kind of RS: one, fully ontology-based and one with enriching the user profile vector with ontology in vector space model (VSM) (proposed method). Afterward the results will be compared with the simple VSM based RS. Given results show that the proposed RS presents the highest performance

*Index Terms*—Recommender system, ontology, data mining, vector space model, wordnet, online forums.

## I. INTRODUCTION

In recent decades among the millions of websites, online discussion forums have always been considered by many of people from all around the world. Online forum is a place for discussing around a specified topic. Content of forums is composed of main groups, sub-groups, discussion topics and posts so that each forum has several main and sub groups which are classified by their content type. In forums there is a role that users are not allowed to post at an irrelevant place or do duplicate posting. Information overload [1] starts in forum when the number of discussions proportion to the number of categories is much. In this case, navigation and access to the information will be hard to the user so that it is so difficult for user to find his desired discussion among the thousands of discussions.

Our initial investigation on 65 of users of Digitalpoint forums who were participated in more that 20 discussions showed that just 6 percent of them will participate in a discussion which is created in a date after the user first entrance to the forum. This means that users have not been successful to find their desired discussion topics among the last topics. Such matter lead to the big problem in discussion forums called "duplicate posts" [2]. Indeed user after not finding his desired topic attempts to create a new discussion or pursue that in a parallel topics and this result in data redundancy problem at forum.

In such circumstances, employment of a suitable IR system, can improve the accessibility to the discussions for users, but these systems performance is not satisfactory when we deal with the high amount of information. On the other hand due to the existence of user profiles in these environments, there is a good potential for improving the retrieval quality which can enables the IR systems to do customization on the results according to the user interests. This result can affects two procedures: one, search process among the discussions (Information Filtering) and another recommending discussions to the users(Recommender System).

Our main goal on this paper is improvement of dynamism of online forums environments and growth of discussions productivity by providing user contributions. The result of our work can be useful for online forum software industry or even news or articles websites.

Look at more general, IR systems and consequently search engines which are presented in the last decades, assume that user is able to express his needs by set of words and bring that to the IR system [3] and thus they don't present a solution for users who are not able to express their needs in the words format. Although information filtering systems presents the users needs according to his interests [4], but these systems are not also able to solve this problem. For solving that, recommender system are proposed which are able to suggest those items to the user so that user has liked them before but he had not been able to express it [5].

Presently, not only RS has become the interest of many researches but also it has been employed in many applications such as e-stores for suggesting books, movies, music, news, etc to the users. The most important problem of RS is problem called cold-start. It means that there is not possible to suggest an item to a user who is new to the system. In this context, RS employ many techniques for making the best suggestion for the users. Most researches divide RS into three different categories: Collaborating Filtering (CF) methods which suggest topics to the users by considering the interest of users who have the highest similarity to the desired user [6]. Content-based filtering which consider user profiles in the past and present the related topics to user based on his profile content. Likewise Hybrid RS employs both of them. Many researches already done show that the collaborating methods are relatively prosper [7] but not in all cases. For instance in the case that the number of users proportion to the number of items is negligible, CF techniques are not able to provide good suggestions [8]. For solving this problem content-based RS could be employed which are rooted in IR techniques.

Traditional content-based RS usually employ the user profile and item content to compute the similarity of them by comparing the user and item keyword vectors based on VSM. In some other techniques such as ontology-based RS, an intermediate ontology is employed for comparing the similarity between user and item profiles. Each of these methods has their own disadvantages. In the first method, there is a semantic gap between user profiles and item profiles so that system is rather sensitive to keywords within the user and item profiles and the second method which both profiles are presented regarding to the existing ontology is vulnerable in case the ontology has low accuracy or quality. In this case, if ontology quality was low, many of concepts may be lost during the comparing process.

Our proposed method is based on vector space model so that the user profiles will be enriched by the constructed ontologies. Likewise ontologies will be learned from domain corpus by combination of text mining and NLP techniques and a lexical database like Wordnet. Our goal is solving both two mentioned problems. It means that the proposed method not only decrease the semantic gap between user and item profiles but also its sensitivity against the ontology quality become less than other methods. Moreover in the case that high quality ontology is not available, it can present more acceptable suggestions.

## II. RELATED WORKS

Related works which will be discussed in the following paragraphs are divided into three subjects of: application of ontology in RS, application of ontology in IR systems and learning ontology from the text.

### A. Ontology-Based Recommender System

One of the earliest researches on presenting the user profiles and documents by ontology in RS backs to the research done by Savia and Jokela [9]. They employed Meta-data for describing taxonomy-based documents. In [10],[11] successful results from employment of hierarchical taxonomy of concepts existing in user profiles for solving of Heterogeneous content problem is reported. Also Ontological Conceptual modeling is employed in Quickstep system [12] which uses four-level ontology and a hybrid RS for suggesting articles to the researchers. In this method, papers are presented as normalized vectors based on weight of concepts. Semantic relations between users and items profiles are calculated by semantic relations of common interests of user and items. Likewise in CoMet project [13] the comparison of user and item profiles is done based on finding the largest branch of the tree shared between user profile and document. In [14] Shoval and colleagues present a new method for calculating the similarity between user and item profiles by comparing them with ontology. In [15] Maidel and colleagues proved that using of taxonomy-based conceptual ontology presents better results from Non-taxonomy-based methods. Likewise in [16] some other methods for computing of similarity between user profiles and items profiles are presented well.

### B. Ontology-Based Information Retrieval

The idea of employment of ontologies in IR systems backs to the first semantic search engines like RDQL,RQL and SPARQL which unstructured information of documents were stored in forms of conceptual ontology and then the search was performed by using Boolean methods [17]. Despite the lack of documents ranking in the mentioned methods, a new method was proposed in SEAL portal [18] which was able to do ranking on the retrieved documents based on the given query. However, there was no evaluation criterion for comparing that method to the others. Rocha, et al. [19] tried to expand queries in a desired ontology and compare the query based on ontology by calculating the similarity between query and results. Nevertheless due to the additive amount of information being generated and non-existence of a standard method for ranking, the Boolean based methods are being useless. Therefore the next idea of semantic search in IR systems is concentrated on keywords. One of applications of ontologies could be in query expansion in order to eliminate the ambiguity of queries. This makes the system to better understand the user orders. In [20] a method is presented that its aim is enriching the user query based on ontology which is constructed by wikipedia. Likewise, in [17] a method is proposed for employment of domain ontology for knowledge-based retrieval.

### C. Ontology Learning From the Text

Many works is done in this area. Gupta and his colleagues [21] proposed a method for extracting a subset of a specific domain from wordnet by the domain corpus which its goal was developing the sub-wordnet for NLP. Khan and Luo [22] constructed their ontology with top-bottom method so that firstly a hierarchical structure constructed by the clustering techniques and the related concepts put into the clusters. Afterward, a specific concept is allocated to a related cluster with the same topic in the tree by bottom-top mechanism. Likewise In [23] domain ontology is constructed by re-using of a bigger ontology. This method is based on using of lexical databases and domain corpus. Indeed the main goal of this work is improving the ontology by standard terminology and vocabulary databases. In another work, Xu et al. [24] related terms to the specific domain are gathered and then relations among them are discovered by text mining and thus the ontology is constructed by this method. Also Farhoodi et al. employed Persian wikipedia to construct the ontology by considering the four level relations: page titles, keywords, related topics and category for discovering the relations among the concepts. In Amini and Abolhasani work [25] a new method is proposed for constructing the semi-ontology by NLP, Statistics and IR techniques in computer science domain.

## III. RESEARCH METHODOLOGY

As seen in the Fig. 1, the implementation components are consisting of Data-set, pre-processing , keyword extracting for building the user , items profiles and three type of ontologies which are constructed by three methods for using as an input in two different recommender system : one, ontology-based recommender system and other the proposed method. Moreover user and item profiles will be used as an input for simple content-based filtering method using VSM. The mentioned components will be discussed in the following sections.

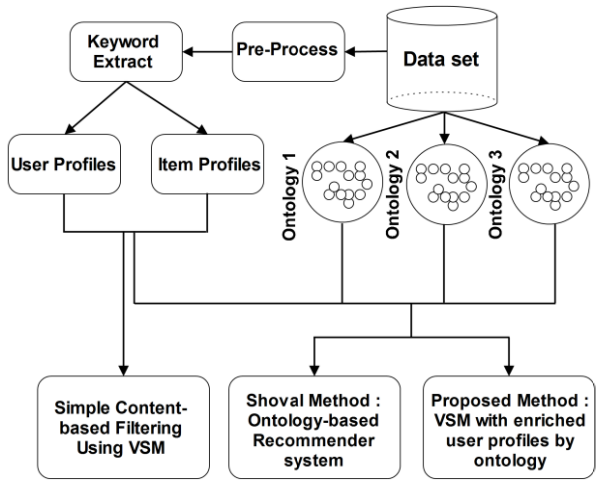

Fig. 1. Research methodology overview.

### A. Data Set

Data set is gathered from one of the famous online forums Digitalpoint forums. Whereas the concentrate of the paper is solving the dispersion problem of existing discussions in a topic, we just extract one group of topics. For this purpose the discussion with topic of entertainment was crawled and saved as HTML documents. Then the post of users extracted from documents by regular expression techniques. Therefore we had 881 discussions which are consist of 24291 posts and 1963 users. The average of user participation in discussions is 4.22 and the maximum count of a user participation in discussions is 234.

### B. Pre-Process

In this section the Bag of Words (BOW) from data set will be constructed. For this purpose, at first the illegal characters, symbols, before and after spaces, common words in forums (e.g. thread, quote …), punctuations, stop-words (e.g. am, is, are, as) will be removed from the documents. Then all of the words convert to lowercase mode and then the high reputation words are eliminated. Then the remained words convert to an array, afterward the porter algorithm [26] will execute on each row of the array (e.g. movie converts to movi). Finally non-repetitive words will be determined with their frequency of them. At final process, our bag of words includes 13.027 words and after execution of porter algorithm has 10.038 words.

### C. User and Item Profiles

For building the user and item profiles, for each item the non-repetitive words of that item with the frequency of that word store in database. For user profile, the number of user posts will be considered addition to its frequency as its weight. For instance if user participated in a discussion for 3 times and a word A exists 5 times in the discussion, the weight of 15 will allocated to word A in user profile.

### D. Ontology Construction

The ontologies is constructed by three ways. First is by Khan and Luo method [22] by hypernyms relation in wordnet (Ontology1) and another by nouns existing in a description of words (gloss words) in wordnet (Ontology2). And the last ontology is constructed by a similar method like Gupta, et, al. method [21] by discovering the relations between the words by wordnet (Ontology3).

Architecture of ontology 1 and 2 construction is illustrated in Fig. 2. As it can be seen, at first, data set is entered in to OntoGen [27] software as an input. Then at first the BOW is created and then all of keywords are weighted by TFIDF method. Then the Latent semantic indexing (LSI) [28] technique is employed to discover concepts with similar semantics relations. With this technique, the synonyms words will be determined. Afterward K-mean clustering will be used for finding the similar groups of discussions which have the highest amount of similar words. K value is optional and it can be selected by evaluating different values of k in order to find the best clustering which its clusters have the highest difference to each other. Next the three most important keywords of each cluster will be determined as the cluster topics. Thereafter in the pre-processing section, the repeated concepts in the higher levels will be removed and then will be enriched by two methods: One by hyponyms relation in wordnet (Ontology 1) and one by extracting of nouns in gloss words (Ontology 2). For this purpose we employ Stanford POS Tagger [29] which is able to extract the nouns from the given gloss.

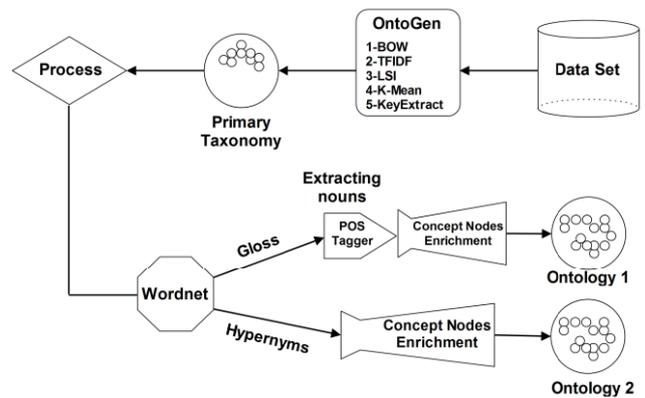

Fig. 2. Architecture of ontology 1 and 2 construction.

In terms of ontology 3, at first the synonyms of each word in BOW will be extracted from wordnet and if the synonym exists in the BOW it will be stored in the brothers table. Likewise same level hypernyms of word will be stored as the words' fathers and a top level hypernyms of word will be stored as the word's grandfathers. For brother, father and grandfather selection there is a criterion of minimum frequency of 10 in BOW.

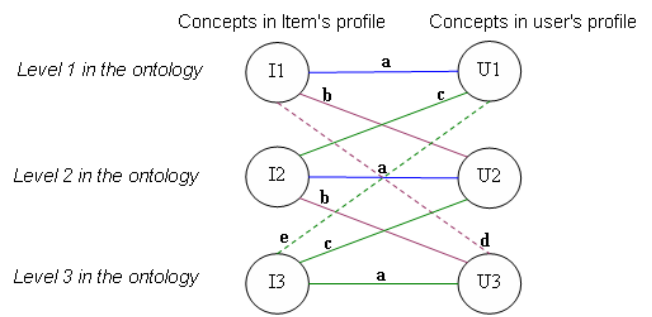

Fig. 3. Hierarchical similarity measure in shoval method [14].

*E. Shoval Recommender System*

Shoval, et, al., [14] presented a new method for calculating similarity of user profiles and item profiles by an intermediate ontology. The ontology used in their work was a three level ontology of IPTC News codes[1]. Since our constructed ontologies are three level it can be possible to evaluate our ontologies in shoval method too. In Shoval method a new method based on weighting of relations between three levels is presented. Fig. 3 shows these three types of relations. Regarding the Fig. 3 shoval coefficients are defined this way [14] : a: $I1=U1, I2=U2, I3=U3$(e.g. both item and user profiles include 'sport') , b:$I1=U2, I2=U3$(e.g. item concept is 'sport', while user concept is 'football') , c:$I2=U1, I3=U2$ (e.g. item concept is 'basketball' while user concept is 'sport'), d: $I1=U3$ (e.g. item concept is 'sport' while user concept is 'Mondeal games'). e: $I3=U1$(e.g. item concept is 'Euro league' while user concept is 'sport'). Then for computing of similarity of user and item profiles the following formula will be used:

$$IS = \frac{\sum_{i \in Z} N_i \cdot S_i}{\sum_{j \in U} N_j} \quad (1)$$

where: *Z*: number of concepts in item's profile, *U*: number of concepts in user's profile, *i*: index of the concepts in item's profile, *j*: index of the concepts in user's profile, *Si*: score of similarity (*a, b, c, d, e*), *Ni*: number of clicks on the concept.

*F. Proposed Recommender System*

In our proposed method, we enrich the user profiles vector in VSM [30] by ontology instead of presenting profiles by ontology. The main goal of our method is decreasing the angle of between user profile and item profile vectors by enriching the user profile by ontology.

Suppose that a user profile set is shown by *U* and Item profiles set is shown by *I*. So, inverse frequency of a term $k_i$ in BOW is calculated as the following formula:

$$idf_i = \log \frac{|I|}{df_i} \quad (2)$$

So that $df_i$ is equal to the items containing the $k_i$ and |*I*| is equal to the total number of items (discussions). Now if we show the term frequency of the $k_i$ in $I_j$ with $tf_{ij}$ we have:

$$TFIDF_{i,j} = tf_{i,j} \times idf_i \quad (3)$$

Afterward, we calculate the TFIDF of each term, and then vector of each user profiles and item profiles will be constructed based on their included terms. These vectors have the same length, so the similarity of these profiles can be calculated by the measurement of cosinus of these vectors devided by normal vectors of them as the following formula:

[1] http://iptc.org/NewsCodes

$$SC(U, I) = \frac{\sum_{1}^{t} TFIDF_U \times TFIDF_I}{\sqrt{\sum_{1}^{t} TFIDF_U^2} \times \sqrt{\sum_{1}^{t} TFIDF_I^2}} \quad (4)$$

The above formula is the main formula which is employed in our simple content-based recommender system. But now we want to enrich the user profile vectors by ontology. So we should add the vector of user profiles three vectors of brothers, fathers and grandfathers of terms existing in the user profiles. For this purpose we first extract the brothers of each terms in user profiles and if extracted brother didn't exists in the user profile we add its calculated TFIDF to the new brother vector. Due to the same length of main vector and its brother we can add these two vectors to build a new enriched user profile vector. But before that, likewise the brother vector, we do the same process for fathers and grandfathers and build the father and grandfather vector of user profiles with the same method. Finally we have four vector that their length are same: main user profile vector ($TFIDF_U$), user profile brother vector ($TFIDF_B$), user profile fathers vector ($TFIDF_F$) and user profiles grandfather vector ($TFIDF_{GF}$). Then the enriched user profile vector will be shown by the following relation:

$$\vec{TFIDF}_{UO} = \vec{TFIDF}_U + \alpha \vec{TFIDF}_b + \beta \vec{TFIDF}_f + \gamma \vec{TFIDF}_{gf} \quad (5)$$

if we show the terms set in user profiles with *K* and $k_i$ is one of terms in K in $U_j$ so that $k_i \in K$ and *B* is brothers set existing in first level of ontology and F is fathers set existing in second level of ontology and GF is grandfathers set existing in third level ontology, then 'b' represents the brothers set of $k_i$ in $U_j$, 'f' represents the fathers set of $k_i$ in $U_j$ and 'gf' represents the grandfathers set of $k_i$ in $U_j$. So that $b \subset B$, $f \subset F, gf \subset GF$ and $b \notin K$, $f \notin K$, $gf \notin K$.

Now for calculating the similarity of user profile and item profiles (SCO) we use the enriched user profile vector instead of old user profile vector in formula(4). So we have:

$$SCO(U, I) = \frac{\sum_{1}^{t} TFIDF_{UO} \times TFIDF_I}{\sqrt{\sum_{1}^{t} TFIDF_{UO}^2} \times \sqrt{\sum_{1}^{t} TFIDF_I^2}} \quad (6)$$

IV. EXPERIMENTS

Before doing the experiments we need to calculate the best coefficients in formula (1) and formula (5). The coefficients in formula(1) already has been calculated by Maidel, et, al. [15] in a real recommender system in ePaper project by doing survey on 57 users in a 4 days period and the best coefficients are calculated as: $a=1, b=0.8, c=0.4, d=0, e=02$. In terms of coefficients of formula (5) in our proposed method, we evaluated many coefficients on a 56 users with our proposed method and ontology 2 and then calculated *F*1 by comparing the generated recommends by user interests existing in data set. The best *F*1 was obtained by $\alpha = 0.8$ ، $\beta = 0.4$ و $\gamma = 0.2$ which will be used in our experiments.

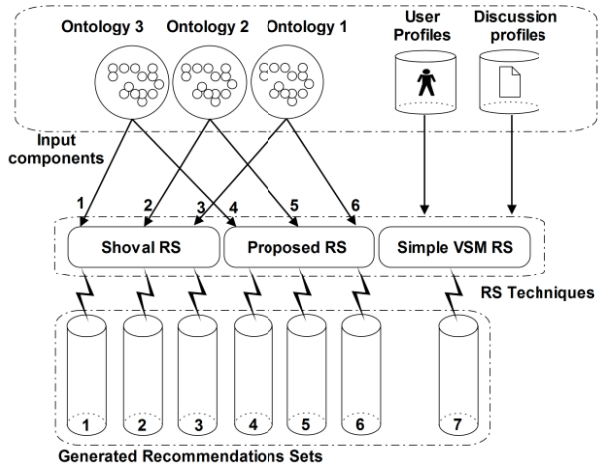

Fig. 4. Experiment process.

Regarding to the Fig. 1, we have three kinds of recommender system and three types of ontologies. One of our recommender systems doesn't use any of ontologies. So, three types of ontologies will be applied in two kinds of recommender system, one with shoval method and one with our proposed method. So, six experiments should be done for generating recommendations. In order to compare results with the simple content-based recommender system which doesn't use ontology we do another experiments on this method too. So seven experiments will be executed and seven recommendations set will be generated at the end of our experiments. Finally these seven recommendation sets will be compared with users' interests existing in our data set for evaluating the methods. The evaluating of these seven experiments will be discussed fully in the next sections.

## V. EVALUATION

There are several ways for evaluation of recommender systems, however according to availability of implicit user interests of users in our data set (participation of user in a discussion will be regarded as his favorite to that discussion) we can use precision and recall measures which are used widely in IR systems. For this purpose at first the list of Top $N$ recommends ($N<=$Total Number of discussions) will be extracted from the generated recommendations in experiments phase and then will be compared to favorites discussions of 56 users who have at least 10 posts in their profile. Due to the high amount of required computing, we just did our evaluation on 35 random $N$.

Since by increasing the number of irrelevant recommendations to the user, precision decreases and increasing the number of recommendation of discussions that must offered to user but not offered, cause the recall to be reduced, we use the F-measure for normalizing the performance evaluation.

The given results are shown in Table I which shows the average $F1$ obtained via the experiments.

Regarding the overall look at Table I, it is clear that our proposed method depending on the employed ontology can improve the recommendation performance from approximately 2 to 10%, however, fully ontology-based methods like shoval overall performance is equal or a little lower than simple VSM RS. It may be due to the quality of ontologies employed in our experiments so that is previously mentioned in [17] that ontology-based methods are sensitive against the quality of ontologies. Moreover since our ontology construction method was automatically, it can not be compared with ontologies which are constructed or revised by a domain expert.

TABLE I: AVG. $F1$ OBTAINED IN TOP $N$ RECOMMENDATION

| Rank | Experiment | Average $F1$ |
|---|---|---|
| 1 | Exp.5: Proposed RS with Ontology2 | 35.8 |
| 2 | Exp.4: Proposed RS with Ontology1 | 32.4 |
| 3 | Exp.6: Proposed RS with Ontology3 | 28.5 |
| 4 | Exp.7: Simple VSM RS | 26.8 |
| 5 | Exp.3: Shoval RS with Ontology3 | 25.9 |
| 6 | Exp.1: Shoval RS with Ontology1 | 24.6 |
| 7 | Exp.2: Shoval RS with Ontology2 | 24.5 |

Comparing the Given F1 in Table I, we can concluded that the best results is presented by ontology 2 and then ontology 1 .the only difference of these ontologies is related to their type of consumption from wordnet. The first one uses the extracted nouns from gloss words in wordnet and the second one uses the hypernyms in wordnet to enrich the concept nodes. So it is obvious that using the nouns existing in gloss, leads to better performance in comparison to the hypernyms. This result is already mentioned in Navigli and Velardi work [31] so that they proved that using of gloss words in wordnet for query expansion, presents more precision in comparison of hypernyms relations.

Comparing the given results for ontology 1, 2 3 which use two different methods, we can conclude that although using wordnet can improve the quality of ontology, the data mining techniques such as clustering can do more in our proposed method.

## VI. CONCLUSION AND FUTURE WORKS

In this paper the performance of three different recommender techniques based on three type of ontologies which was evaluated in order to improve the quality of recommendations in online forums. For this purpose, we used one groups of digitalpoint forums as a data set for evaluating our results. Our proposed method was based on enriching the user profile by a three level ontology which was constructed from our domain corpus. Also we used wordnet for enchaining the concept nodes or extracting the parent-child relations. With regards to the given results we can conclude that using constructed ontology from domain corpus with using text mining or wordnet can improve the performance of Simple VSM based techniques. But in terms of ontology-based techniques like shoval work, due to the sensitivity of these ontology-based methods to the ontology, the results will not be interesting. For these methods we can use updated and high quality ontologies which are made by made or revised by domain experts instead of automatic ontologies. Moreover regarding the results, we can see that the quality of recommendation presented by ontology-based methods is very near to the simple VSM based method. So in the large scale projects that computing time is important and also no high quality is available, ontology-based methods like shoval which their required computation volume is lower that VMS based method can be the best option.

We used wordnet as our lexical database. Recent researches show that in some cases, other sources like wikipedia can have better impacts. Moreover we used the implicit feedbacks of users. In [12], [32] it is proved that the explicit feedbacks of users can improve the recommendations. So one of the future work can be designing the prototype of the proposed method and get the explicit feedbacks of users directly and compare the results to the given results.

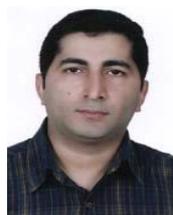

**Hadi Fanaee Tork** currently is a PhD student of computer science at Laboratory of Artificial intelligence and decision support (LIAAD), University of Porto, Portugal. Before that he was research fellow at IEETA, University of Aveiro, Portugal for one year. He holds MSc degree in information technology from Shiraz University, Iran and BSc in Industrial metallurgy engineering from Ferdowsi University of Mashhad, Iran. He worked more than 10 years in different companies , R&D centers and governmental sections as a programmer and researcher Presently, his main research interests are  Event detection, Spatial data ranking, Spatio-temporal data mining, recommender systems and text mining.

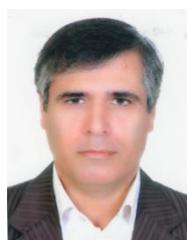

**Mehran Yazdi** received his B.Sc. degree in Digital Communication Systems from the Department of Electrical Engineering, Shiraz University, Shiraz, Iran in 1992 and M.Sc. and Ph.D. degrees in Digital Vision and Image Processing from the Department of Electrical Engineering, Laval University, Quebec, Canada in 1996 and 2003 respectively. He is currently Associate Professor at the Department of Communication and Electronic Engineering, Shiraz University, Iran. He conducted several projects in the area of hyperspectral image compression and denoising, CT metal artifact reduction, video compression and distributed video systems. His major research interests are in the field of image/video processing, remote sensing, multidimensional signal processing, E-learning and medical image analysis.